\documentclass[preprint,conference]{IEEEtran}

\usepackage{configuration}

\acresetall

\begin{document}

\bstctlcite{IEEEexample:BSTcontrol}

\AddEverypageHook{%
  \ifnumequal{\value{page}}{1}{%
    \tikz[remember picture,overlay]{%
      \node[
        draw,
        minimum width=1.03\textwidth,
        text width=1.02\textwidth,
        font=\scriptsize,
        align=center,
        inner sep=4pt
      ]
      at ($(current page header area) - (0,5pt)$)
      {%
        © 2026 IEEE. Personal use of this material is permitted. Permission from IEEE must be obtained for all other uses, in any current or future media, including reprinting/republishing this material for advertising or promotional purposes, creating new collective works, for resale or redistribution to servers or lists, or reuse of any copyrighted component of this work in other works. DOI: 10.1109/XXXXXX.XXXX.XXXXXXXX.
      };
    }%
  }{}%
}

\title{
    Performance Analysis of ISAC Slicing with Sensing-Assisted
    Communication and Communication-Assisted Sensing
}
\author{
    \IEEEauthorblockN{Daniel \yale{Dante} Campos and Jacek Kibi\l{}da\IEEEauthorrefmark{2}}
    \vspace{-0.1in}
    \\\IEEEauthorblockA{\IEEEauthorrefmark{2}\textit{Commonwealth Cyber Initiative}, \textit{Virginia Tech}, USA, emails: 
    \{dcamposfsa, jkibilda\}@vt.edu}
    \vspace{-0.3in}
}

\maketitle

\begin{abstract}
Integrated Sensing and Communication (ISAC) will allow 6G networks to provide both sensing and communication services through shared infrastructure and spectrum resources.
However, network slicing that views these services as distinct slices with distinct objectives and performance requirements will inevitably miss the emergent value of integrating them. 
By considering Sensing-Assisted Communication (SAC) and Communication-Assisted Sensing (CAS), which create interdependencies between services, we can observe trade-offs in resources dimensioned to support each slice and, ultimately, the ISAC slicing performance.
As such, this paper proposes a performance analysis of an ISAC slicing framework that accounts for resource allocation and sensor density, as well as SAC and CAS, dimensioned to achieve slice-specific Service-Level Agreement (SLA) satisfaction.
Numerical simulations, across different bandwidth allocations, sensor deployments, and demand scenarios, show that ISAC slicing performance is fundamentally constrained by sensing resolution, shaped by demand, and scaled by sensor deployment.
These findings provide insights for the design of future ISAC slicing mechanisms.
\end{abstract}


\section{Introduction}%
\label{sec:introduction}

One of the major innovations behind future 6G networks is \ac{ISAC}, which is based on the idea of co-designing wireless systems to enable simultaneous provisioning of communication and sensing services.
While much of the innovation in terms of integration and performance analysis is currently occurring at the physical and signal-processing layers~\cite{ofdm_sensing}, the system-level view remains somewhat limited, positing the existence of two types of slices: %
\ac{CaaS}, and \ac{SaaS}~\cite{awesome_isac_survey, dong2022sensing, shafi2025industrial}.
Viewed as distinct slices with distinct objectives and performance requirements, sensing and communication will inevitably compete for resources~\cite{li2025joint, li2023toward}.

In this article, we make the case that, unlike conventional slicing, \ac{ISAC} slicing introduces interdependencies between \ac{SaaS} and \ac{CaaS} slices.
Communication performance may benefit from sensing via \emph{\ac{SAC}}, leveraging sensing information to enhance control decisions, such as beamforming based on position estimates of served \acp{UE}~\cite{awesome_isac_survey, zhou2024temporal}.
Simultaneously, sensing accuracy and coverage may benefit from \emph{\ac{CAS}}, which includes providing communication links for sensors in bi- and multistatic sensing modes to share data~\cite{huawei_isac_specs, chih2025cooperative}.

\begin{figure}[t]
    \centering
    \includegraphics[width=\linewidth]{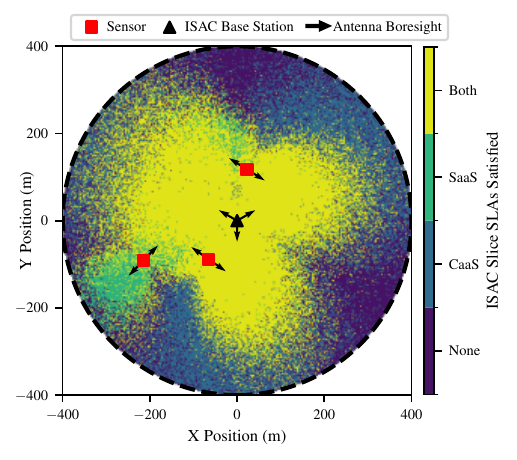}
    \caption{Sample \ac{SLA} satisfaction coverage for one \ac{SaaS} slice and one \ac{CaaS} slice in a single-cell \ac{ISAC} slicing scenario with three deployed sensors performing bistatic sensing.}
    \label{fig:0_scenario}
    \vspace{-0.3in}
\end{figure}

These interdependencies are likely to become particularly poignant in scenarios with high target densities. %
Sensing in such scenarios requires additional bandwidth to maintain sufficient sensing resolution, ultimately limiting the number of distinguishable targets~\cite{ofdm_sensing,signal_fundamentals}. %
Beyond resource allocation, sensor density itself becomes a critical design variable for \ac{ISAC} networks supporting bi- and multistatic sensing~\cite{chih2025cooperative, fatima2024optimal}.
By impacting %
sensing distances, sensor density  %
directly affects sensing accuracy and coverage~\cite{chih2025cooperative, fatima2024optimal, li2023toward}.
Consequently, dimensioning resources to satisfy \ac{SaaS} and \ac{CaaS} \acp{SLA} presents a challenging question that requires accounting for both demand and network-specific factors. %
In this article, we evaluate these factors together in an ISAC slicing scenario and characterize their impact on sensing, communication, and, ultimately, network-level slicing performance.
\Fig{0_scenario} illustrates a snapshot of network-level slice \ac{SLA} satisfaction across a cell provisioning \ac{ISAC}.

Recent work has investigated resource allocation, cooperative sensing, and ISAC slicing, but these aspects are studied independently and evaluated with different performance metrics.
The works~\cite{li2025joint, li2023toward} optimize beamforming, sensor-target assignment, and sensing coverage in bi- and multistatic cooperative sensing deployments, evaluating performance through \ac{SNR} under fixed sensing topologies.
Sensor deployment has also been investigated as a sensing-design problem.
In~\cite{fatima2024optimal}, the authors optimize sensor placement to improve sensing accuracy, while~\cite{chih2025cooperative} surveys deployment, selection, and coordination mechanisms for cooperative \ac{ISAC}.
Still, these works do not address slicing and resource allocation.
On the other hand, studies on these two topics, such as \cite{dong2022sensing,guan2024slicing,hossain2023ai}, optimize bandwidth, power, or deployment decisions under sensing and communication requirements, but, %
importantly, treat sensing and communication as independent.
Recent works have separately demonstrated the benefits of \ac{SAC} for communication through sensing-assisted beamforming~\cite{zhou2024temporal} and highlighted the communication overhead of \ac{CAS} in cooperative sensing~\cite{sensing_data_rate}.
Nevertheless, these interdependencies are not analyzed within the context of \ac{ISAC} slicing.
To the best of our knowledge, this work is the first to evaluate, in a bistatic \ac{ISAC} slicing scenario, how \ac{SAC}, \ac{CAS}, and sensing deployment impact on sensing, communication, and network-level slicing performance.

Our contributions can be summarized as follows:
\begin{itemize}
    \item We propose an \ac{ISAC} slicing framework that jointly models the impact of resource allocation, sensor deployment, \ac{SAC}, and \ac{CAS}, linking network configuration to sensing, communication, and slicing performance.
    \item We develop a publicly available simulation tool and evaluate trends and trade-offs between \ac{SaaS} and \ac{CaaS} with \ac{SAC} and \ac{CAS} across different sensor deployment densities, bandwidth allocations, and demand scenarios.
    \item We find that \ac{ISAC} slicing performance is: \1 constrained by resolution, which impacts both slices and defines the \ac{SAC} trade-off; \2 shaped by demand, especially from \ac{CaaS} slices; and finally \3 scaled by sensor deployment, significantly improving \ac{SaaS} slice performance. %
\end{itemize}

\section{Network Model}
\label{sec:network_model}

This section describes the key network and channel modeling parameters and assumptions, with a focus on the \ac{SNR}, which is the basis for describing the link budget and associated performance of both sensing and communication services.

\subsection{Network}
We consider a single \ac{UMi} in a vehicular scenario where \iac{ISAC} \ac{BS}, denoted by $M$, provides communication and sensing services to \ac{SaaS} and \ac{CaaS} slices.
The cell is divided into a set of $\Kc$ non-overlapping sectors.
Without loss of generality, we assume that each sector is equipped with \iac{ULA} of $A_{\bs}$ antenna elements.
Within the cell range $R$, there %
is a set of vehicles $\Vc$,
of which $\Uc \subseteq \Vc$ are active \acp{UE} attached to the \ac{BS} and served by the \ac{CaaS} slice; and $\Tc \subseteq \Vc$ are known targets tracked by the \ac{SaaS} slice.
Since \ac{OFDM} continues to be the waveform of choice for the upcoming releases of 6G~\cite{shafi2025industrial}, we assume that the \ac{BS} implements \ac{OFDM}-based \ac{ISAC}, as in, e.g.,~\cite{wei20225g, ofdm_sensing}, leveraging time and frequency resources.
Without loss of generality, we assume a frame of $T$ time slots, and an operating bandwidth of $B$ Hz around a center frequency $f_{c}$.
To support the sensing service, a set of $\Nc$ sensors is deployed at fixed and known locations within the cell range, each equipped with \iac{ULA} of $A_{\sen}$ antenna elements, collecting sensing data, and transmitting it in-band to the \ac{BS} to support both slices.

\subsection{Sensing Link Budget}
We assume a bistatic sensing scenario in which the \ac{ISAC} \ac{BS} acts as the sensing transmitter, while a sensor $n \in \Nc$ acts as the sensing receiver.
As such, we assume that the two paths of the cascaded sensing channel, $D_v$, from the \ac{BS} to $v\in\Vc$, and $D_{v,n}$, from $v$ to $n\in\Nc$, are both 
\ac{LOS}, following~\cite{signal_fundamentals,li2025joint}.
Thereby, we obtain the bandwidth-normalized sensing \ac{SNR}, as follows
\begin{equation}
    \gamma_{\sens}(v,n) =
    \frac{
        P_{\bs} G_{\procgain}(v) G_{1}(\bs,v) G_{1}(n,v) \lambda^{2} \sigma_{\mathrm{rcs}}(v,n) 
    }{
        (4\pi)^{3}  (D_{v}  D_{v,n})^{\alpha}  N_{0} \varrho_v \varrho_{v,n}
    }, 
\end{equation}
where $P_{\bs}$ is the per-beam transmit power uniformly allocated between downlink beams, $G_{\procgain}(v)$ is the sensing processing gain (defined as below), $G_{1}(\bs, v)$ is the directional gain of the transmitter \ac{BS} antenna, $G_{1}(n,v)$ is the directional gain of the receiver sensor antenna, $\lambda$ is the wavelength of $f_{c}$, $\sigma_{\mathrm{rcs}}(v,n)$ is the \ac{RCS} (calculated as in \cite[Table 7.9.2.1-4]{3gpp_channel_model}), $N_{0}$ is the noise power spectral density, and $\varrho_v,\varrho_{v,n}\sim \operatorname{log}\Nc(0,\sigma_{\mathrm{F}}^2)$ are the large-scale fading on the two paths with variance of $\sigma_{\mathrm{F}}^2$. %
For simplicity, we consider $G_{1}(\bs, v)$ and $G_{1}(n,v)$ as the maximum directional gain (calculated as in~\cite[Table 7.3-1]{3gpp_channel_model}).
We generalize the path-loss exponent $\alpha \ge 2$ to model scenarios with additional attenuation. 
Finally, the processing gain $G_{\procgain}(v)$ is obtained for \ac{OFDM}-based estimation following~\cite{ofdm_sensing}:
\begin{equation}
    \label{eq:processing_gain}
    G_{\procgain}(v) = A_{\sen} N_{\beam}  \min(l_{v} \tau, \tau_{\mathrm{CPI}}),
\end{equation}
where $N_{\beam} = \frac{B_{\sens}}{\Delta_f}$ is the number of sensing subcarriers with $B_{\sens}$ the sensing bandwidth and $\Delta_f$ the subcarrier spacing, $l_{v}$ is the number of beam time slots allocated to $v$, $\tau$ is the number of \ac{OFDM} symbols in a slot, and $\tau_{\mathrm{CPI}}$ is the number of \ac{OFDM} symbols in the \ac{CPI}.

\subsection{Communication Link Budget}

\noindent\textbf{Downlink --}
On both downlink and uplink communication channels, we consider the path-loss following 3GPP's Urban Microcell Street Canyon \ac{LOS} model~\cite{3gpp_channel_model}.
Given $n \in \Nc$, the sensor assigned to track $u\in\Uc$, we write the bandwidth-normalized downlink communication \ac{SNR} as 
\begin{equation}
    \gamma_{\ue}(u,n) =
    \frac{
        P_{\bs}   G_{\beam}(u,n)  G_{\ue}
    }{
        L(D_{u}) N_0 \varrho_u
    },
\end{equation}
where $L(D_{u})$ denotes path-loss at distance $D_{u}$ (calculated as in \cite[Table 7.4.1-1]{3gpp_channel_model}), $G_{\ue}$ is the receiver \ac{UE} single-antenna isotropic gain, and $\varrho_u\sim \operatorname{log}\Nc(0,\sigma_{\mathrm{F}}^2)$ is the large-scale fading. $G_{\beam}(u,n)$, the sensing-based beamforming gain, is defined following~\cite{beam_misalignment}:
\begin{equation}
    G_{\beam}(u,n) =
    \begin{cases}
        G_{1}(\bs, u)  10^{-\frac{3}{10}\big(\frac{2 \sigma_{\beam}(u,n) }{\varphi_{\mathrm{A}}}\big)^2} \text{if } \sigma_{\beam}(u,n) \le \Theta_u \\
        G_{2} \qquad\qquad\qquad\qquad\quad\;\; \text{otherwise}, %
    \end{cases}
\end{equation}
where $\varphi_{\mathrm{A}}=\pi - 2\arccos(\frac{1.391\lambda}{\pi A_{\bs}d})$ is the broadside \ac{HPBW} of the \ac{ISAC} \ac{BS} with $d$ the antenna element spacing~\cite[Table 6.2]{antenna_design}, $G_2=(A_{\bs}\sin^{2}(\frac{3\pi}{2A_{\bs}}))^{-1}$ is the side lobe beamforming gain, $\Theta_u=\frac{\varphi_{\mathrm{A}}}{2} \sqrt{\frac{10}{3}\log_{10}(\frac{G_{1}(\bs, u)}{G_{2}})}$ is the mainlobe width of the beam serving $u$, and $\sigma_{\beam}(u,n)$ is the beam misalignment error (as defined in \Sec{isac_framework}).

\noindent\textbf{Uplink --}
In the uplink, which the sensor uses to report the sensing measurements to the network, a sensor $n\in\Nc$ positioned at a distance $D_n$ of the \ac{BS} experiences the bandwidth-normalized \ac{SNR}
\begin{equation}
    \gamma_{\sen}(n) =
    \frac{
        P_{\sen} G_{1}(n,\bs) G_{1}(\bs, n)
    }{
        L(D_{n})  N_0 \varrho_n
    },
\end{equation}
where $P_{\sen}$ is the sensor's transmit power, $G_{1}(n, \bs)$ is the directional gain of the transmitter sensor antenna, $G_{1}(\bs, n)$ is the directional gain of the receiver {BS} antenna, and $\varrho_n\sim \operatorname{log}\Nc(0,\sigma_{\mathrm{F}}^2)$ is the large-scale fading.
Since the \ac{BS} and sensor positions are fixed, we assume maximum directional gain.

\section{\ac{ISAC} Slicing Framework}
\label{sec:isac_framework}

In this section, we describe our \ac{ISAC} slicing framework, comprising bandwidth and beam time slot allocation for \ac{SaaS} and \ac{CaaS} slices, target-sensor assignment, \ac{SAC} and \ac{CAS} interdependencies, and performance metrics.

\subsection{ISAC Slicing}

\noindent\textbf{\ac{SaaS} and \ac{CaaS} --}
We consider the \ac{SaaS} slice to support sensing-based applications, such as environment monitoring and \ac{ISAC}-aided \ac{V2X}, by providing a localization service that tracks detected vehicles~\cite{awesome_isac_survey}. %
To evaluate its performance, we choose the position estimation error metric.
We assume that a single position estimate per frame is sufficient to accurately track vehicle motion.
Also, we consider the impact of sensing resolution at the network level, limiting the number of resolvable vehicles.
Accordingly, we describe the \ac{SaaS} \ac{SLA} as a max. position estimation error $Q_{\sens}$.
The \ac{CaaS} slice, on the other hand, provides a %
\ac{V2X} \ac{GBR} communication service~\cite[Table 5.7.4-1]{3gpp_5qi}.
We then define its \ac{SLA} as a minimum throughput per frame $Q_{\comm}$.

\noindent\textbf{\ac{SAC} and \ac{CAS} --}
The sensing functions of the \ac{ISAC} \ac{BS} can be leveraged not only to satisfy the \ac{SaaS} \ac{SLA}, but also to improve communication performance and \ac{CaaS} \ac{SLA} satisfaction through \ac{SAC}.
We model \ac{SAC} as the opportunistic use of sensing bandwidth for a sensor $n \in \Nc$ to track a \ac{UE} $u \in \Uc$ and reduce its beam misalignment $\sigma_{\beam}(u,n) = \arctan(\frac{\sigma_{\position}(u,n)}{D_u})$, for $\sigma_{\position}(u,n)$ the position estimation error.
However, \ac{UE} and target position estimations can be used efficiently for \ac{SAC} and \ac{SaaS} only if they are continuously transmitted %
from sensors to the \ac{ISAC} \ac{BS}, \ie if the \ac{CAS} demand is satisfied.
Further in this section, we model the position estimation error and the impact of unsatisfied \ac{CAS} demand on tracking success.

\subsection{Resources Allocation}
\noindent
\textbf{Bandwidth Allocation --}
Assuming \ac{FDD} operation, the bandwidth $B$ is divided between downlink $B_{\dl}$, to sense and serve vehicles, and uplink $B_{\ulink}$, for sensors to report measurements.
We further split downlink bandwidth between $B_{\comm}$, communication bandwidth associated with the \ac{CaaS} slice, and $B_{\sens}$, sensing bandwidth associated with the \ac{SaaS} slice.
The sensing bandwidth is used solely for sensing and is entirely allocated to each simultaneous beam.
With frequency reuse, each \ac{UE} $u \in \Uc$ is allocated a downlink communication bandwidth $b_{u} \le B_{\comm}$, while each sensor is allocated an uplink communication bandwidth $b_{n} \le B_{\ulink}$.

\noindent
\textbf{Beam Allocation --}
Each sector $k \in \Kc$ can simultaneously sense/serve $A_{\bs}$ vehicles and $A_{\bs}$ sensors in multi-beam operation by distributing a budget of $A_{\bs}T$ downlink and $A_{\bs}T$ uplink beam time slots.
Assuming at most one beam per device, each vehicle $v \in \Vc$ and sensor $n \in \Nc$ are allocated  $l_{v} \le T$ and $l_{n} \le T$ slots, respectively.

\begin{remark}
To manage communication interference with frequency reuse, as in~\cite{dong2022sensing}, we assume resources are allocated orthogonally.
On the downlink, if a \ac{UE} $u \in \Uc$ has a set of overlapping \acp{UE} $\Uc_{u} \subseteq \Uc$ that might be served simultaneously ($l_u+\sum_{i \in \Uc_{u}} l_i> T$), we consider they are assigned non-overlapping channels ($b_{u} + \sum_{i \in \Uc_{u}} b_{i} \le B_{\comm}$).
Likewise, on the uplink, if a sensor $n \in \Nc$ has an overlapping set $\Nc_{n} \subseteq \Nc$ with $l_n+\sum_{j \in \Nc_{n}} l_j> T$, we assume $b_{n} + \sum_{j \in \Nc_{n}} b_{j} \le B_{\ulink}$.
\end{remark}

\subsection{Service Performance}

\noindent\textbf{Communication Performance --}
The uplink data rate $\psi_{\sen}(n)$ of a sensor $n \in \mathcal{N}$ and downlink data rate $\psi_{\ue}(u,n)$ of a \ac{UE} $u \in \mathcal{U}$ tracked by $n$ are obtained as
\begin{equation}
    \label{eq:thr_sensor}
    \psi_{\sen}(n) =
    \frac{l_n}{T} b_{n}  \log_{2}(1 + \gamma_{\sen}(n)),
\end{equation}
\begin{equation}
    \label{eq:thr_ue}
    \psi_{\ue}(u, n) =
    \frac{l_u}{T} b_{u}  \log_{2}(1 + \gamma_{U}(u,n)).
\end{equation}

\noindent\textbf{Sensing Accuracy --}
We obtain the accuracy of estimating the position of a vehicle $v \in \mathcal{V}$ tracked by a sensor $n\in\Nc$ as
\begin{equation}
    \label{eq:position_error}
    \sigma_{\position}(v,n) = \sqrt{
        \sigma_{\range}(v,n)^2 + (\sigma_{\angleerr}(v,n) D_{v,n})^2
    },
\end{equation}
where $\sigma_{\range}(v,n)$ and $\sigma_{\angleerr}(v,n)$ denote the errors in range and angle estimation relative to $n$, respectively.
For both, we assume efficient, unbiased estimators, and approximate their average estimation error as the square root of the \ac{CRLB}~\cite{signal_fundamentals}, which can be written as
\begin{equation}
    \label{eq:range_error}
    \sigma_{\range}(v,n) =
    \frac{c}{
        4 \pi B_{\sens} \sqrt{2\gamma_{\sens}(v,n)}
    },
\end{equation}
\begin{equation}
    \sigma_{\angleerr}(v,n) =
    \frac{
        \lambda \sqrt{6}
    }{
        2\pi d \max (\epsilon, |\cos(\theta_{v,n})|) \sqrt{\gamma_{\sens}(v,n)A_{\sen}(A_{\sen}^2-1)}
    },
\end{equation}
where $c$ is the speed of light, $\theta_{v,n}$ is the relative angle between $v$ and the boresight of the \ac{ULA} of sensor $n$.
Since the term $\cos(\theta_{v,n})$ in $\sigma_{\angleerr}(v,n)$ %
comes from \ac{CRLB}'s far field assumption and does not apply as $\theta_{v,n} \rightarrow$ 90º, we lower-bound it by $\epsilon$.

\noindent
\textbf{Sensing Capacity --}
Tracking is only feasible if sensors meet their \ac{CAS} demand, and the sensing resolution is fine enough to distinguish each vehicle.
Given $I(v,n) \in \{0,1\}$, the vehicle-sensor assignment, where each vehicle is assigned one sensor, we write the rate of successfully tracked vehicles as
\begin{equation} %
    \rho =
    \frac{ \min(|\Vc|, Y)}{|\mathcal{V}|^2}
    \sum_{n \in \Nc} \min \bigg(\bigg\lfloor\frac{\psi_{\sen}(n)}{Z}\bigg\rfloor, \sum_{v \in \Vc} I(v,n) \bigg),
\end{equation}
where $Y = \frac{\pi R^2}{\Delta_{\mathrm{c}}}$ approximates the maximum number of resolvable vehicles for \ac{OFDM}-based sensing~\cite{ofdm_sensing} and $\Delta_\mathrm{c} = \frac{c}{2 B_{\sens}} R \frac{\lambda}{A_{\sen} d}$,
is the resolution cell area~\cite{signal_fundamentals} for $R$ assuming the worst-case cross-range resolution.

\subsection{Slice Performance Satisfaction}

We evaluate the impact of resource allocation and target assignment using slice- and network-level aggregated metrics.
On the \ac{SaaS} slice, we use the mean position estimation error
\begin{equation}
    \overline{\sigma_{\tar}} =
    ( 1-\rho) R
    +  
    \frac{\rho}{|\Tc|}
    \sum_{t \in \Tc}
    \sum_{n \in \Nc}
    I(t,n)\sigma_{\position}(t,n),
\end{equation}
where $R$ is the error experienced by unsuccessfully tracked targets.
On the \ac{CaaS} slice, we calculate the mean throughput
\begin{equation}
    \overline{\psi_{\ue}}
    =
   \frac{(1-\rho)}{|\Uc|}
    \sum_{u \in \Uc}
    \psi_{\ue_{2}}(u)
    +
    \frac{\rho}{|\Uc|}
    \sum_{u \in \Uc}
    \sum_{n \in \Nc}
    I(u,n)\psi_{\ue}(u,n),
\end{equation}
where $\psi_{\ue_{2}}(u)$ is the throughput achieved by $u \in \Uc$ if it is not successfully tracked, thus experiencing the side lobe beamforming gain $G_{2}$.
Finally, we evaluate the \ac{ISAC} slicing satisfaction at the network level by adapting the \ac{SSR} metric from~\cite{guan2024slicing}:
\begin{align}
    \overline{Q}
    &=
    \nonumber
    \frac{\rho W_{\sens}}{|\Tc|}
    \sum_{t \in \mathcal{T}}
    \mathds{1}_{
        \big\{
        \sum\limits_{n \in \mathcal{N}}
        I(t,n) \sigma_{\position}(t,n)
        \le Q_{\sens}
        \big\}
    }
    \\
    \nonumber
    &+
    \frac{\rho W_{\comm}}{|\Uc|}
    \bigg(
        \sum_{u \in \Uc}
        \mathds{1}_{
            \big\{
                \sum\limits_{n \in \mathcal{N}}
                I(u,n)\psi_{U}(u,n)
                \ge Q_{\comm}
            \big\}
        }
    \bigg)
    \\
    &+
    \frac{(1-\rho)W_{\comm}}{|\Uc|}
    \bigg(
        \sum_{u \in \Uc}
        \mathds{1}_{
            \big\{
                \psi_{\ue_{2}}(u)
                \ge Q_{\comm} 
            \big\}
        }
    \bigg),
\end{align}
where $W_{\sens}$ and $W_{\comm}$ are normalized weights defined by the network operator to prioritize slices.
The first term is the weighted \ac{SaaS} \ac{SSR}, while the second and third are the weighted \ac{CaaS} \ac{SSR} for tracked and untracked \acp{UE}, respectively.

\section{Simulated Environment}
\label{sec:simulated_environment}

In this section, we detail the parameters and allocation strategies implemented in our \ac{ISAC} slicing simulation\footnote{\url{github.com/DanteCampos/2026-ISAC-DJ}}.
Table \ref{tab:simulation_parameters} summarizes simulation parameters, which we adopt from the mmWave \ac{OFDM}-based \ac{ISAC} work~\cite{ofdm_sensing} and the 3GPP \ac{UMi} model~\cite{3gpp_channel_model}.

\begin{table}[t]
\caption{Simulation Parameters}
\label{tab:simulation_parameters}
\centering
\footnotesize
\begin{tabular}{p{4.2cm} p{0.9cm} p{2.4cm}}

\toprule
\textbf{Parameter} &
\textbf{Symbol} &
\textbf{Value} \\
\midrule

Cell Radius &
$R$ &
\unit[400]{m} \\

Number of Sectors &
$|\Kc|$ &
3~\cite{3gpp_channel_model} \\

Carrier Frequency &
$f_c$ &
\unit[28]{GHz}~\cite{ofdm_sensing} \\

Wavelength &
$\lambda$ &
$c/f_{c}$=\unit[10.71]{mm}~\cite{ofdm_sensing} \\

Effective System Bandwidth &
$B$ &
\unit[60]{MHz}~\cite{ofdm_sensing} \\

Subcarrier Spacing &
$\Delta_f$ &
\unit[120]{kHz}~\cite{ofdm_sensing} \\

Slots per frame &
$T$ &
80
~\cite{3gpp_frame_structure}
\\

\ac{OFDM} Symbols per Slot &
$\tau$ &
14
~\cite{3gpp_frame_structure}
\\

\ac{OFDM} Symbols in the \ac{CPI} &
$\tau_{\mathrm{CPI}}$ &
512~\cite{ofdm_sensing} \\

\ac{ULA} Antenna Elements &
$A_{\bs}, A_{\sen}$ &
16~\cite{ofdm_sensing} \\

\ac{ULA} Antenna Spacing &
$d$ &
$\lambda /2$=\unit[5.35]{mm}~\cite{ofdm_sensing} \\

\ac{UE} Single Isotropic Antenna Gain &
$G_{\ue}$ &
\unit[0]{dBi}~\cite{beam_misalignment}\\

Sensing Path Loss Exponent &
$\alpha$ &
2.4 \\

BS Transmit Power per Beam &
$P_{\bs}$ &
\unit[23]{dBm}~\cite{3gpp_channel_model} \\

Sensor Transmit Power &
$P_{\sen}$ &
\unit[23]{dBm} \\

Noise Power Spectral Density &
$N_{0}$ &
\unit[-174+9]{dBm/Hz}~\cite{3gpp_channel_model} \\

Large-Scale Fading Variance &
$\sigma_{\mathrm{F}}^2$ &
\unit[4]{dB}~\cite{3gpp_channel_model} \\

Sensing Data Generation per Vehicle &
$Z$ &
\unit[29.72]{kbps} \cite{sensing_data_rate} \\

Cosine Lower Bound Correction &
$\epsilon$ &
$\cos$(75º) \\

Slice Weights &
$W_{\comm}, W_{\sens}$ &
0.5 \\

\ac{SaaS} Max Pos. Est. Error Req. &
$Q_{\sens}$ &
\unit[0.1]{m}~\cite{huawei_isac_specs}\\

\ac{CaaS} Min Throughput Req. &
$Q_{\comm}$ &
\unit[24]{Mbps}~\cite{v2x_overview}\\

\bottomrule
\end{tabular}
\end{table}

\subsection{Network Configuration}

We consider a single cell with the \ac{ISAC} \ac{BS} at the origin.
Sensors, targets, and \acp{UE} are independently distributed within the cell area according to a homogeneous \acl{PPP} of arbitrary densities, while sensor orientations are uniform at random.
The \ac{ISAC} \ac{BS} is equipped with 3 sectorized \acp{ULA} transmitting at \unit[35]{dBm}, following~\cite{3gpp_channel_model}, which we uniformly distribute across 16 downlink beams as \unit[23]{dBm/beam}.
We consider a single isotropic antenna \ac{UE}, and a \ac{ULA} sensor.
Sensors report back their collected data once per \ac{OFDM} frame, in a message we assume is similar in size to a Collection Perception Message~\cite{sensing_data_rate}.
We define two slices with equal weights: one \ac{SaaS} slice provides sensing to Urban Environment Reconstruction \cite{huawei_isac_specs}, while one \ac{CaaS} slice provides communication to \ac{V2X} applications \cite{v2x_overview}.

\subsection{Resource Allocation}
For each slice, we design resource allocation to guarantee orthogonality in the time or frequency domains.
To isolate the impact of resource allocation, we consider upper-bound sensor assignments with minimum position estimation error via exhaustive best-selection.
We leave the study of practical assignment algorithms for future work.

On the downlink, we allocate beam time slots to each vehicle $v \in \Vc$ in a Round-Robin manner, limiting $l_v \le \tau_{\mathrm{CPI}}/\tau$ if $v \notin \Uc$ to allocate the remaining slots to \acp{UE}.
Since sensing and communication operate on separate bandwidths, we allocate the entire communication bandwidth to \acp{UE}: $b_{u} \gets B_{\comm}$ $\forall u \in \Uc$.
We then identify the set $\Uc_u$ for each \ac{UE} $u \in \Uc$, containing $u$ and all \acp{UE} whose beams it overlaps with.
If $\Uc_u$ overlaps in the time domain ($\sum_{j\in\Uc_u} l_{j} > T$), we enforce frequency-domain orthogonality by updating: $b_i \gets \min(b_i , \frac{B_{\comm}}{|\Uc_{u}|})$ $\forall i \in \Uc_u$.

In the uplink, we do the opposite.
Due to the maximum directional gains between the fixed sensors and the \ac{ISAC} \ac{BS}, and to the large subcarrier spacing, we are able to satisfy the \ac{CAS} demand of all sensors by reusing \unit[1]{\ac{PRB}}.
Therefore, we fix this allocation to dedicate the remaining \acp{PRB} to downlink communication and sensing, while enforcing orthogonality across beam time slots.
Sensors start with $l_{n} \gets T$ $\forall n \in \Nc$, then we check the set of sensors $\Nc_{n}$ with $n$ and the beam it overlaps with.
If $\sum_{j\in\Nc_n} l_{j} > T$, we update $l_j \gets \min(l_j , \frac{T}{|\Nc_{n}|})$ $\forall j \in \Nc_n$.

\section{Evaluation}
\label{sec:evaluation}

In this section, we evaluate the performance of \ac{ISAC} slicing across different downlink sensing-communication bandwidth splits, sensor deployments, and demand scenarios, defined by target and \ac{UE} densities.
Without loss of generality, we consider $\Tc \cap \Uc = \emptyset$.
Since targets have a lower impact on beam slot allocation due to the $\tau_{\mathrm{CPI}}$ bound, we consider target densities higher than \ac{UE} densities.
For short, we refer to the bandwidth split as the \ac{NSB} $= \frac{B_{\sens}}{B_{\dl}}$, with $B_{\dl} = B_{\sens} + B_{\comm}$, and abbreviate targets, \acp{UE}, and sensors as T, U, and S, respectively.
Results are averaged over 1000 independent realizations for statistical significance.

\subsection{\ac{ISAC} Slicing Satisfaction}

\begin{figure}[t]
    \centering
    \includegraphics[width=\linewidth]{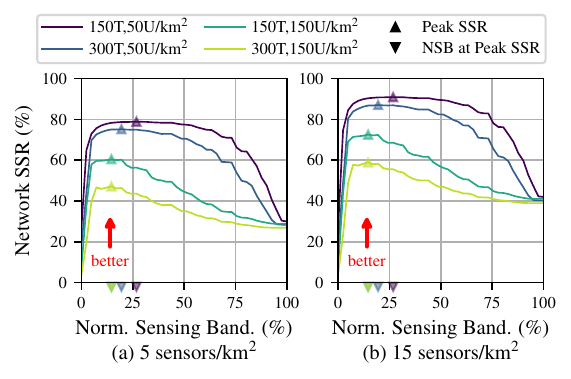}
    \caption{\ac{ISAC} Slicing performance at network-level.}
    \label{fig:3_slice_performance}
\end{figure}

\Fig{3_slice_performance} displays the network \ac{SSR} as a function of \ac{NSB} under different sensor deployments and demand scenarios.
We observe the \ac{SaaS}-\ac{CaaS} trade-off: at \unit[100\%]{\ac{NSB}}, no \ac{CaaS} \ac{SLA} is satisfied, whereas at \unit[0\%]{\ac{NSB}}, no \ac{SaaS} \ac{SLA} is satisfied.
The \ac{NSB} at the highest network \ac{SSR} is therefore a compromise between the two slices.
The best-performing bandwidth split is primarily determined by demand, particularly \ac{CaaS} demand, since we do not constrain \ac{UE} beam time slot allocation.
Consequently, increasing demand shifts the best-performing \ac{NSB} %
towards a larger share of the bandwidth for communication.

Increasing the sensor density has little effect on the trade-off itself, scaling the achievable \ac{SSR} without significantly altering the curve's shape. This behavior is illustrated in \Fig{3_slice_performance}a for the deployment with \unit[5]{sensors/km$^2$}.
The low-demand scenario (150T, 50U/km$^2$) achieves a peak network \ac{SSR} of 79\% at \unit[27\%]{\ac{NSB}} and remains within 2 percent points of it over the \ac{NSB} interval [10\%, 54\%].
In contrast, the high-demand scenario (300T, 150U/km$^2$) reaches a lower peak \ac{SSR} of 47\% at \unit[15\%]{\ac{NSB}}, with a narrower interval of [7\%, 20\%].
This result indicates that \ac{ISAC} slicing becomes increasingly sensitive to bandwidth allocation as demand increases.
Furthermore, \ac{CaaS} demand has a stronger impact than \ac{SaaS} demand.
Increasing only the \ac{UE} density (150T, 150U/km$^2$) reduces the peak \ac{SSR} by 18 points, whereas increasing only the target density (300T, 50U/km$^2$) reduces it by 4 points.
Increasing the sensor density to \unit[15]{sensors/km$^2$}, shown in \Fig{3_slice_performance}b, provides an almost uniform increase of 12 points across all demand scenarios.
Overall, bandwidth allocations that satisfy the \ac{SaaS}-\ac{CaaS} trade-off are constrained by \ac{CaaS} demand, whereas achievable network performance scales with sensor density.
To better understand these effects, we next analyze the \ac{CaaS} and \ac{SaaS} slices individually.

\subsection{\ac{CaaS} Performance and \ac{SAC} Trade-Off}

\begin{figure}[t]
    \centering
    \includegraphics[width=\linewidth]{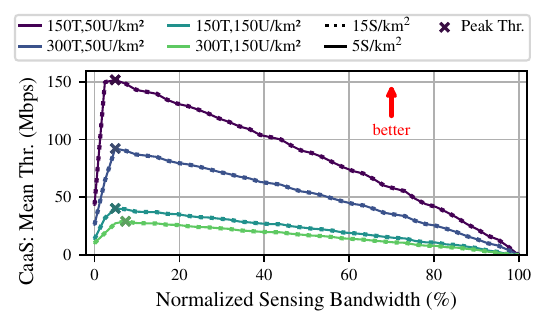}
    \caption{\ac{CaaS} performance and \ac{SAC} trade-off.}
    \label{fig:2_communication_performance}
\end{figure}

\Fig{2_communication_performance} evaluates the \ac{CaaS} performance.
We observe the \ac{SAC} trade-off: at \unit[100\%]{\ac{NSB}}, no bandwidth remains for communication, whereas at \unit[0\%]{\ac{NSB}}, throughput is degraded by beam misalignment.
Peak performance occurs at the lowest \ac{NSB} with enough resolution to successfully track all vehicles.
Beyond this point, the \ac{SNR} gain from improved beamforming is outweighed by the loss of communication bandwidth.
Since beamforming requires substantially lower positioning accuracy than the \ac{SaaS} slice, communication performance is not limited by sensing accuracy, but by resolution and demand.
As a result, \ac{CaaS} performance is largely insensitive to sensor density.

In the low-demand scenario (150T, 50U/km$^2$), the mean throughput peaks at \unit[152]{Mbps} with \unit[5\%]{\ac{NSB}}.
In the high-demand scenario (300T, 150U/km$^2$), this peak decreases to \unit[29]{Mbps} and requires a higher sensing allocation of \unit[7\%]{\ac{NSB}} to resolve all vehicles.
Consistent with the network \ac{SSR} results, \ac{CaaS} demand dominates the achievable performance.
Increasing only the \ac{UE} density (150T, 150U/km$^2$) reduces throughput by approximately \unit[73]{\%} relative to the low-demand scenario.
In contrast, increasing only the target density (300T, 50U/km$^2$) reduces throughput by approximately \unit[39]{\%}.

\subsection{\ac{SaaS} Performance}

\begin{figure}[t]
    \centering
    \includegraphics[width=\linewidth]{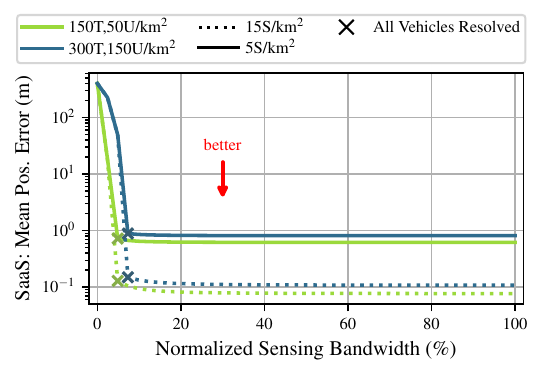}
    \caption{\ac{SaaS} performance of tracking \ac{SaaS} targets.}
    \label{fig:1_sensing_performance}
\end{figure}

\Fig{1_sensing_performance} depicts the \ac{SaaS} performance.
We observe that sensing performance is first limited by resolution, then by sensor density.
Once all vehicles are resolved, increasing the \ac{NSB} yields only marginal improvements because the additional receiver noise offsets the processing gain.
In contrast, sensor density has a substantial impact on sensing performance.
Denser deployments reduce the target-sensor distance, improving the link budget and decreasing the cross-range component of the position error, as shown in \Eq{position_error}.
Finally, \ac{SaaS} is less sensitive to demand than \ac{CaaS}, since targets require fewer beam slots to achieve the \ac{CPI}.

These effects are reflected in the position estimation error.
With sparse deployments (5S/km$^2$), the mean error at \unit[100\%]{\ac{NSB}} reduces only slightly from \unit[0.81]{m} in the high-demand scenario (300T, 150U/km$^2$) to \unit[0.62]{m} in the low-demand scenario (150T, 50U/km$^2$).
In contrast, increasing the sensor density to 15S/km$^2$ significantly reduces the error, \unit[0.11]{m} and \unit[0.08]{m} in the high- and low-demand scenarios, respectively.
Therefore, unlike the \ac{CaaS} slice, \ac{SaaS} performance tolerates demand increases and is primarily determined by sensor density once sensing bandwidth is sufficient to resolve all vehicles.

Combined with the communication performance results, this explains the behavior of the network \ac{SSR}.
Both slices are fundamentally constrained by resolution, while the best-performing bandwidth split is dominated by \ac{CaaS} demand.
Consequently, as demand increases, the peak network performance tends to be achieved by the \ac{NSB} with the smallest sensing bandwidth share that still resolves all vehicles.

\section{Conclusions}
\label{sec:conclusion}

In this work, we proposed a model for evaluating the sensing, communication, and slicing performance of an \ac{ISAC} cell with bistatic sensing via a network of sensors.
Our results reveal a hierarchy of design decisions for \ac{ISAC} slicing networks.
First, sufficient sensing bandwidth is allocated to resolve all vehicles, as sensing resolution is the primary bottleneck for both \ac{SaaS} and \ac{CaaS} slices.
Second, the sensor deployment is dimensioned to meet the \ac{SaaS} \ac{SLA} requirements, as this is the main factor determining sensing accuracy.
Finally, the sensing-communication bandwidth split that defines the \ac{SaaS}-\ac{CaaS} trade-off is considered in relation to the demand scenario, with a primary focus on how \ac{CaaS} \ac{SLA} satisfaction degrades.
Future work will focus on designing resource-allocation algorithms to dynamically resolve the observed trade-offs and ensure \ac{SLA} satisfaction for both \ac{SaaS} and \ac{CaaS} slices.

\section*{Acknowledgments}
This work received support from the National Science Foundation under IUCRC WISPER (No. 2412872), and Miageru (No. 2554414) grants, and from the Commonwealth Cyber Initiative (CCI) in Virginia, US, an investment in advancing cyber R\&D, innovation, and workforce development.

\bibliographystyle{IEEEtran}
\bibliography{references}

@IEEEtranBSTCTL{IEEEexample:BSTcontrol,
  CTLuse_forced_etal       = "yes",
  CTLmax_names_forced_etal = "7",
  CTLnames_show_etal       = "7"
}

@article{beam_misalignment,
  title={{Effects of SNR-Dependent Beam Alignment Errors on Millimeter-Wave Cellular Networks}},
  author={Zia, Muhammad Saad and Blough, Douglas M and Weitnauer, Mary Ann},
  //author={Zia, Muhammad Saad and Blough, Douglas M and others},
  journal={IEEE Transactions on Vehicular Technology},
  volume={71},
  number={5},
  pages={5216--5230},
  year={2022},
  publisher={IEEE}
}

@techreport{3gpp_channel_model,
    author={3GPP},
    title = {{5G; Study on Channel Model for Frequencies from 0.5 to 100 GHz (3GPP TR 38.901 v19.3.0)}},
    institution = {3GPP},
    year = {2026}
}

@techreport{3gpp_5qi,
    author={3GPP},
    title = {{5G; System architecture for the 5G System (5GS) (3GPP TS 23.501 v19.7.0)}},
    institution = {3GPP},
    year = {2026}
}

@techreport{3gpp_frame_structure,
    author={3GPP},
    title = {{5G; NR; Physical Channels and Modulation (3GPP TS 38.211 v19.3.0)}},
    institution = {3GPP},
    year = {2026}
}

@book{antenna_design,
  title={{Antenna Theory: Analysis and Design}},
  author={Balanis, Constantine A},
  year={2016},
  publisher={John Wiley \& Sons}
}

@book{signal_fundamentals,
  title     = "Fundamentals of Radar Signal Processing, Third Edition",
  author    = "Richards, Mark",
  publisher = "McGraw-Hill Education",
  edition   =  3,
  month     =  apr,
  year      =  2022,
  address   = "Columbus, OH",
  language  = "en"
}

@article{awesome_isac_survey,
  title={{Integrated Sensing and Communications: Toward Dual-Functional Wireless Networks for 6G and Beyond}},
  author={Liu, Fan and Cui, Yuanhao and Masouros, Christos and Xu, Jie and Han, Tony Xiao and Eldar, Yonina C and Buzzi, Stefano},
  //author={Liu, Fan and others},
  journal={IEEE Journal on Selected Areas in Communications},
  volume={40},
  number={6},
  pages={1728--1767},
  year={2022},
  publisher={IEEE}
}

@article{hossain2023ai,
  title={{AI-Assisted E2E Network Slicing for Integrated Sensing and Communication in 6G Networks}},
  author={Hossain, Mohammad Arif and Xiang, Amanda and Kiani, Abbas and Saboorian, Tony and Kaippallimalil, John and Ansari, Nirwan},
  journal={IEEE Internet of Things Journal},
  volume={11},
  number={6},
  pages={10627--10634},
  year={2023},
  publisher={IEEE}
}

@article{guan2024slicing,
  title={{Slicing-Aware Aerial Networks for Integrated Sensing and Communication: 3D Placement and Adaptive Allocation of Resources}},
  author={Guan, Yingying and Song, Qingyang and Chen, Tao and Qi, Weijing and Guo, Lei and Jamalipour, Abbas},
  journal={IEEE Vehicular Technology Magazine},
  volume={19},
  number={2},
  pages={79--88},
  year={2024},
  publisher={IEEE}
}

@article{ofdm_sensing,
  title={{A Novel Joint Angle-Range-Velocity Estimation Method for MIMO-OFDM ISAC Systems}},
  author={Xiao, Zichao and Liu, Rang and Li, Ming and Liu, Qian and Swindlehurst, A Lee},
  //author={Xiao, Zichao and Liu, Rang and Li, Ming and Liu, Qian and others},
  journal={IEEE Transactions on Signal Processing},
  volume={72},
  pages={3805--3818},
  year={2024},
  publisher={IEEE}
}

@article{v2x_overview,
  title={{An Overview of 3GPP Cellular Vehicle-to-Everything Standards}},
  author={Wang, Xuyu and Mao, Shiwen and Gong, Michelle X},
  journal={GetMobile: Mobile Computing and Communications},
  volume={21},
  number={3},
  pages={19--25},
  year={2017},
  publisher={ACM New York, NY, USA}
}

@techreport{huawei_isac_specs,
  title={{Integrated Sensing and Communication (ISAC) —- From Concept to Practice}},
  author={Bayesteh, Alireza and He, Jia and Chen, Yan and Zhu, Peiying and Ma, Jianglei and Shaban, Ahmed Wagdy and Yu, Ziming and Zhang, Yunhao and Zhou, Zhi and Wang, Guangjian},
  //author={Bayesteh, Alireza and He, Jia and Chen, Yan and others},
  journal={Communications of Huawei Research},
  pages={4--25},
  year={2022}
}

@article{sensing_data_rate,
title = {{Improving Object Selection for Collective Perception Messages Under Congestion}},
journal = {Ad Hoc Networks},
volume = {185},
pages = {104175},
year = {2026},
//issn = {1570-8705},
//doi = {https://doi.org/10.1016/j.adhoc.2026.104175},
//url = {https://www.sciencedirect.com/science/article/pii/S1570870526000417},
author = {Andreia Figueiredo and João Amaral and Pedro Rito and Miguel Luís and Susana Sargento},
//keywords = {Autonomous mobility, Collective perception, Cooperative perception messages, Smart city, Vehicular network},
}

@article{shafi2025industrial,
  title={{Industrial Viewpoints on RAN Technologies for 6G}},
  author={Shafi, Mansoor and Larsson, Erik G and Lin, Xingqin and Panaitopol, Dorin and Parkvall, Stefan and Ronteix-Jacquet, Flavien and Toskala, Antti},
  //author={Shafi, Mansoor and Larsson, Erik G and Lin, Xingqin and others},
  journal={arXiv preprint arXiv:2508.08225},
  year={2025}
}

@article{li2025joint,
  title={{Joint Target Assignment and Resource Allocation for Multi-Base Station Cooperative ISAC in AAV Detection}},
  author={Li, Ruotong and Zhang, Qixun and Ma, Dingyou and Yu, Kan and Huang, Yuzhen},
  //author={Li, Ruotong and others},
  journal={IEEE Transactions on Vehicular Technology},
  volume={74},
  number={5},
  pages={7700--7714},
  year={2025},
  publisher={IEEE}
}

@article{li2023toward,
  title={{Toward Seamless Sensing Coverage for Cellular Multi-Static Integrated Sensing and Communication}},
  author={Li, Ruoguang and Xiao, Zhiqiang and Zeng, Yong},
  //author={Li, Ruoguang and Xiao, Zhiqiang and others},
  journal={IEEE Transactions on Wireless Communications},
  volume={23},
  number={6},
  pages={5363--5376},
  year={2023},
  publisher={IEEE}
}

@article{fatima2024optimal,
  title={{Optimal Placement of the Receivers for Multistatic Target Localization}},
  author={Fatima, Ghania and Stoica, Petre and Aubry, Augusto and De Maio, Antonio and Babu, Prabhu},
  journal={IEEE Transactions on Radar Systems},
  volume={2},
  pages={391--403},
  year={2024},
  publisher={IEEE}
}

@article{chih2025cooperative,
  title={{Cooperative ISAC: An End-to-End Perspective}},
  author={Chih-Lin, I and Han, Zixiang and Xi, Rongyan and Wang, Tianxiong and Zhang, Xiaozhou and Jiang, Tao and Wang, Sen and Zhao, Hanting and Gui, Xin and Jin, Jing and others},
  journal={IEEE Journal of Selected Topics in Electromagnetics, Antennas and Propagation},
  year={2025},
  publisher={IEEE}
}

@article{dong2022sensing,
  title={{Sensing as a Service in 6G Perceptive Networks: A Unified Framework for ISAC Resource Allocation}},
  author={Dong, Fuwang and Liu, Fan and Cui, Yuanhao and Wang, Wei and Han, Kaifeng and Wang, Zhiqin},
  //author={Dong, Fuwang and Liu, Fan and others},
  journal={IEEE Transactions on Wireless Communications},
  volume={22},
  number={5},
  pages={3522--3536},
  year={2022},
  publisher={IEEE}
}

@article{zhou2024temporal,
  title={{Temporal-Assisted Beamforming and Trajectory Prediction in Sensing-Enabled UAV Communications}},
  author={Zhou, Shengcai and Yang, Halvin and Xiang, Luping and Yang, Kun},
  //author={Zhou, Shengcai and Yang, Halvin and others},
  journal={IEEE Transactions on Communications},
  volume={73},
  number={7},
  pages={5408--5419},
  year={2024},
  publisher={IEEE}
}

@article{wei20225g,
  title={{5G PRS-Based Sensing: A Sensing Reference Signal Approach for Joint Sensing and Communication System}},
  author={Wei, Zhiqing and Wang, Yuan and Ma, Liang and Yang, Shaoshi and Feng, Zhiyong and Pan, Chengkang and Zhang, Qixun and Wang, Yajuan and Wu, Huici and Zhang, Ping},
  //author={Wei, Zhiqing and others},
  journal={IEEE Transactions on Vehicular Technology},
  volume={72},
  number={3},
  pages={3250--3263},
  year={2022},
  publisher={IEEE}
}

\end{document}